# The LUVOIR Ultraviolet Multi-Object Spectrograph (LUMOS): Instrument Definition and Design

Kevin France[a,b], Brian Fleming [a,*] Garrett West [c], Stephan R. McCandliss[d], Matthew R. Bolcar[c], Walter Harris[e], Leonidas Moustakas[f], John M. O'Meara[g], Ilaria Pascucci[e], Jane Rigby[c], David Schiminovich[h], Jason Tumlinson[i], Jean-Claude Bouret[j], Christopher J. Evans[k], Miriam Garcia[l]

[a]Laboratory for Atmospheric and Space Physics, University of Colorado, Boulder CO 80309, USA
[b]Center for Astrophysics and Space Astronomy, University of Colorado, Boulder CO 80309, USA
[c]NASA Goddard Space Flight Center, Greenbelt MD 20771, USA
[d]Department of Physics and Astronomy, Johns Hopkins University, Baltimore MD 21218, USA
[e]Lunar and Planetary Laboratory, University of Arizona, Tucson AZ 85721, USA
[f]NASA Jet Propulsion Laboratory, Oak Grove, CA 91109, USA
[g] Department of Chemistry and Physics, Saint Michael's College, Colchester, VT 05439, USA
[h] Department of Astronomy, Columbia University, New York, NY 10027, USA
[i]Space Telescope Science Institute, Baltimore MD 21218, USA
[j] Aix Marseille Univ, CNRS, LAM, Laboratoire d'Astrophysique de Marseille, Marseille, France
[k] UK Astronomy Technology Centre, Royal Observatory, Blackford Hill, Edinburgh, EH9 3HJ, UK
[l] Centro de Astrobiología (CSIC-INTA), Ctra. de Torrejón a Ajalvir, km 4 28850 Torrejón de Ardoz, Spain
[*]NASA Nancy Grace Roman Fellow

## Abstract

The *Large Ultraviolet/Optical/Infrared Surveyor* (*LUVOIR*) is one of four large mission concepts currently undergoing community study for consideration by the 2020 Astronomy and Astrophysics Decadal Survey. *LUVOIR* is being designed to pursue an ambitious program of exoplanetary discovery and characterization, cosmic origins astrophysics, and planetary science. The *LUVOIR* study team is investigating two large telescope apertures (9- and 15-meter primary mirror diameters) and a host of science instruments to carry out the primary mission goals. Many of the exoplanet, cosmic origins, and planetary science goals of *LUVOIR* require high-throughput, imaging spectroscopy at ultraviolet (100 – 400 nm) wavelengths. The LUVOIR Ultraviolet Multi-Object Spectrograph, LUMOS, is being designed to support all of the UV science requirements of *LUVOIR*, from exoplanet host star characterization to tomography of circumgalactic halos to water plumes on outer solar system satellites. LUMOS offers point source and multi-object spectroscopy across the UV bandpass, with multiple resolution modes to support different science goals. The instrument will provide low ($R$ = 8,000 – 18,000) and medium ($R$ = 30,000 – 65,000) resolution modes across the far-ultraviolet (FUV: 100 – 200 nm) and near-ultraviolet (NUV: 200 – 400 nm) windows, and a very low resolution mode ($R$ = 500) for spectroscopic investigations of extremely faint objects in the FUV. Imaging spectroscopy will be accomplished over a 3 × 1.6 arcminute field-of-view by employing holographically-ruled diffraction gratings to control optical aberrations, microshutter arrays (MSA) built on the heritage of the Near Infrared Spectrograph (NIRSpec) on the *James Webb Space Telescope (JWST)*, advanced optical coatings for high-throughput in the FUV, and next generation large-format photon-counting detectors. The spectroscopic capabilities of LUMOS are augmented by an FUV imaging channel (100 – 200nm, 13 milliarcsecond angular resolution, 2 × 2 arcminute field-of-view) that will employ a complement of narrow- and medium-band filters.

The instrument definition, design, and development are being carried out by an instrument study team led by the University of Colorado, Goddard Space Flight Center, and the LUVOIR Science and Technology Definition Team. LUMOS has recently completed a preliminary design in Goddard's Instrument Design Laboratory and is being incorporated into the working *LUVOIR* mission concept. In this proceeding, we describe the instrument requirements for LUMOS, the instrument design, and technology development recommendations to support the hardware required for LUMOS. We present an overview of LUMOS' observing modes and estimated performance curves for effective area, spectral resolution, and imaging performance. Example "LUMOS 100-hour Highlights" observing programs are presented to demonstrate the potential power of *LUVOIR*'s ultraviolet spectroscopic capabilities.

kevin.france@colorado.edu

# 1. INTRODUCTION: ULTRAVIOLET SPECTROSCOPY WITH LUVOIR

NASA's 2013 Astrophysics Roadmap (*Enduring Quests, Daring Visions: NASA Astrophysics in the Next Three Decades*) identified two large mission concepts that would be capable of detecting and characterizing potentially inhabited exoplanets, the Large Ultraviolet/Optical/InfraRed (LUVOIR) Surveyor and the ExoEarth Mapper, a scaled-down version of which has become the Habitable Planet Explorer (HabEx). In 2015, community-led mission studies by NASA's Cosmic Origins Program Analysis Group (COAPG) and Exoplanet Analysis Group (EXOPAG) concluded that both of these missions were meritorious and recommended that reference mission concepts be developed for submission to the 2020 Decadal Survey. While HabEx is focused mainly on exoplanets, the LUVOIR concept is more ambitious in scope, proposing a statistical study of Earth-like planets in the solar neighborhood and a program of ambitious UV/O/IR astrophysics, in the tradition of the Hubble Space Telescope. NASA has convened Science and Technology Definition Teams for LUVOIR and HabEx; mission architecture and instrument design studies for these missions are ongoing, to be completed in 2019.

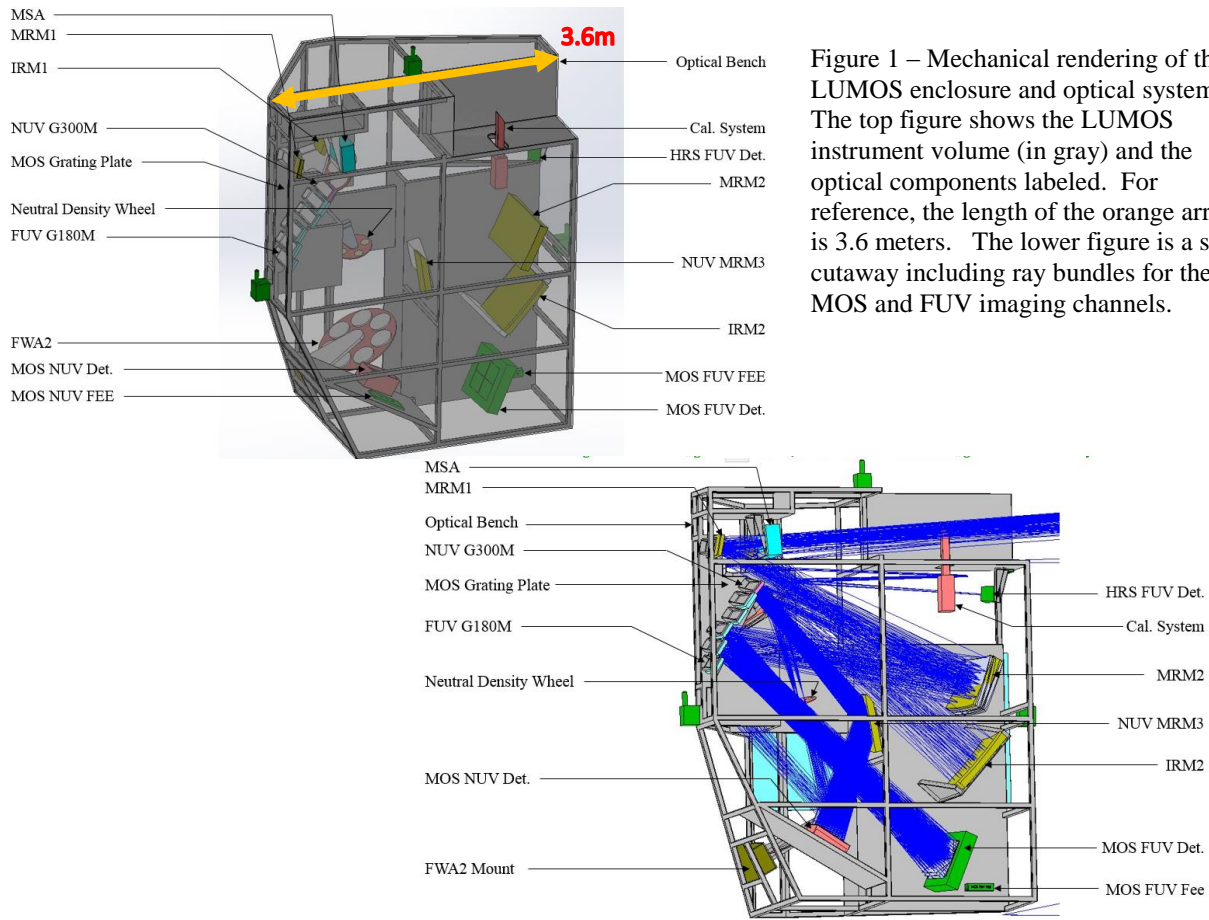

Figure 1 – Mechanical rendering of the LUMOS enclosure and optical system. The top figure shows the LUMOS instrument volume (in gray) and the optical components labeled. For reference, the length of the orange arrow is 3.6 meters. The lower figure is a side cutaway including ray bundles for the MOS and FUV imaging channels.

In their review of future large mission concepts, the COPAG's recommendation for the instrumentation suite for LUVOIR was clear: detection and characterization of rocky planets, wide-field astronomical imaging, and high-throughput ultraviolet imaging and spectroscopy were all deemed essential. The COPAG report to NASA's Science Mission Directorate on flagship missions to be studied by the 2020 Decadal Survey stated "A flagship mission offering … access to the full range of wavelengths covered by HST … is essential to advancing key Cosmic Origins science goals in the 2020s and 2030s."[5]. This conference proceeding describes the work of the LUVOIR Science and Technology Definition Team (STDT) to develop an instrument concept for a broadly capable ultraviolet spectrograph and imager, the LUVOIR Ultraviolet Multi-Object Spectrograph (LUMOS).

*Ultraviolet Spectroscopy and Imaging with LUVOIR* - The LUVOIR STDT, working with a large number of community-led science interest groups, has developed a comprehensive set of science drivers for the mission. A ubiquitous theme that has emerged in the science definition phase has been the study of gas in the cosmos, its relationship to (and evolution with) star and galaxy formation, and how this gas is transferred from one site to another. Understanding the flow of matter and energy from the intergalactic medium (IGM) to the circumgalactic media (CGM), and ultimately into galaxies where it can serve as a reservoir for future generations of star and planet formation, is essentially a challenge in characterizing the ionic, atomic, and molecular gas at each phase in this cycle. LUVOIR will be capable of characterizing the composition and temperature of this material in unprecedented scope and detail; on scales as large as the cosmic web and as small as the atmospheres of planets around other stars.

In the IGM and CGM, the gas is hot and diffuse, best studied in broad atomic hydrogen lines (the Lyman series of H) and metal lines tracing the temperature range of a few hundreds of thousands of degrees to a few millions of degrees (e.g., O VI and Ne VIII) over the past 10 billion years. Once this gas has been accreted into the interstellar medium of galaxies, we observe it in a suite of low-ionization metals and atomic species (e.g., Si II, Mg II, CII, C I, Lyman series lines of H and D). This material coalesces into protostars and their surrounding protoplanetary disks where hot gas lines (e.g., C IV) serve as tracers of mass-accretion and the dominant molecular species ($H_2$ and CO) are used to characterize the planet-forming environment. After these planets grow and emerge from their natal disks, their long-term atmospheric stability can be probed in exospheric species such as H I, O I, and C II while the atmospheric markers for life on rocky planets in their respective habitable zones (HZs) are thought to be $O_2$, $O_3$, $CO_2$, and $CH_4$. The common denominator for all of these tracers is that the strongest emission and absorption lines – therefore the highest information content for understanding the physical conditions in these objects – reside at ultraviolet wavelengths, roughly 100 – 400nm. The LUMOS instrument is designed to make revolutionary observational contributions to all of the disciplines that call for high-resolution spectroscopy, multi-object spectroscopy, and imaging in the ultraviolet bandpass.

## 2. THE LUVOIR ULTRAVIOLET MULTI-OBJECT SPECTROGRAPH: LUMOS

The LUMOS concept is designed to execute the broad LUVOIR science program described in Section 1; the LUVOIR science portfolio is ambitious, and as such several instrument modes will be required to fulfill the objectives of the mission. LUMOS is a highly multiplexed ultraviolet spectrograph, with medium and low-resolution multi-object imaging spectroscopy and FUV imaging modes. LUMOS can be thought of as an analog to the successful *HST*-STIS instrument, with two orders-of-magnitude higher efficiency, multi-object capability, and a wide-field multi-band imaging channel[2]. Example science studies that one can imagine as 'LUVOIR Cycle 1 Large Programs' are presented in Section 4 to present the breadth of science and instrumental capability offered by LUMOS.

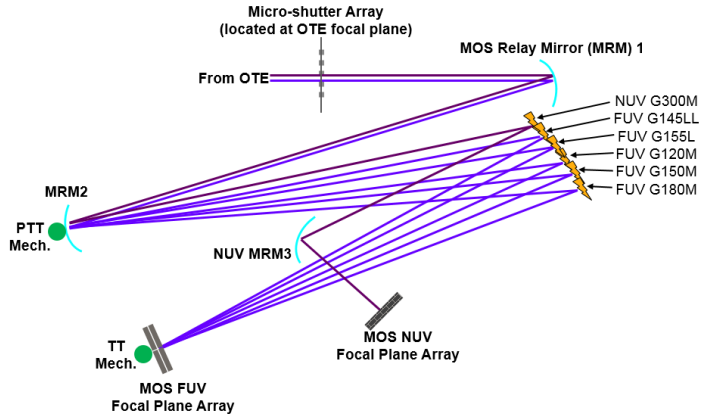

Figure 2 – Schematic raytrace of the LUMOS multi-object spectrograph mode. Light from the optical telescope element (Architecture A, 15-meter primary diameter) enters through the microshutter array before being relayed to the appropriate grating (shown at the upper right). The grating disperses and focuses the light onto the FUV detector bank (a 2 x 2 grid of MCP detectors) or NUV detector bank (a 3 x 7 grid of δ-doped CMOS devices).

LUMOS is fed by the LUVOIR Optical Telescope Assembly (OTE; see Bolcar et al. 2017 – this volume). The LUVOIR study is considering two variants of the OTE, an 'Architecture A' that features a segmented 15.1-meter primary mirror and an 'Architecture B' with a 9.2-meter segmented primary mirror. LUMOS was designed to be compatible with the *f*/20 Architecture A design, and that is the version presented here. A second version of LUMOS (LUMOS-modB) will be designed specifically to work with the smaller telescope in Fall 2017. Coupling the high instrumental throughput with a factor of ~40× gain in collecting area over HST ($[15m / 2.4m]^2 = 40$), LUMOS can reach to limiting fluxes of order 100 – 1000 times fainter than currently possible with the HST-COS and –STIS instruments.

LUMOS is built into a temperature-controlled enclosure (Figure 1), maintained at 280K (warmer than the rest of the observatory) to prevent outgassing species to settle onto to any of the UV optics or FUV microchannel plate (MCP) detectors. The NUV complementary metal-oxide-semiconductor (CMOS) detectors are the only component of LUMOS that are operated cold, with a nominal operating temperature of 170K.

### 2.1 LUMOS Multi-Object Imaging Spectroscopy Modes

Figure 2 shows the optical layout of the LUMOS multi-object imaging spectroscopy channel. The entrance aperture for the spectrograph is a 3 x 2 grid of microshutter arrays[41] (MSAs), located 2.28′ × 1.98′ from the center of the optical focus of the OTE. The MSA arrays build on the heritage of NIRSPEC on *JWST*[42], with 6 individual arrays of 480 x 840 shutters where each shutter has a 100µm x 200µm pitch. The MSA grid defines the field-of-view for multi-object spectroscopy, 3′ × 1.6′ for the FUV modes and 1.3′ × 1.6′ for the NUV mode. Each shutter has a projected angular size of 0.136″ × 0.068″, with a 63% unvignetted open area (0.108″ × 0.054″). In both modes, light passing through the MSA is folded into the spectrograph by a nominally fixed (with fine focus control for calibration) convex biconic optic, the Multi-Object Spectrograph (MOS) Relay Mirror 1 (MRM1; Figures 1 and 2). A second aberration correcting toroidal steering mirror (MRM2) with both piston and tip-tilt control directs the beam to one of six fixed gratings, and at this point the light path becomes wavelength-dependent.

*FUV modes* - The FUV multi-object spectroscopy modes include a set of medium, low, and very low spectral resolution settings that all maintain sub-MSA shutter imaging performance, essentially creating an array of long-slits that can be used for point-source spectroscopy or extended source imaging spectroscopy. The gratings are on fixed mounts along one side of the LUMOS optical bench and are selected with the appropriate motion of the MRM2 transfer optic (Figure 1). The medium resolution gratings provide bandpass averaged $R$ = 20,000 – 70,000 (depending on grating setting and field location) imaging spectroscopy over most of the 3′ × 1.6′ field-of-view. The grating names are chosen in analogy with the HST naming convention, where 'G' refers to a first order grating, the middle three numbers are roughly the central wavelength in nanometers, and the final letter ('M', 'L', or 'LL') denotes the spectral resolution of the mode (Medium, Low, or Low-Low).

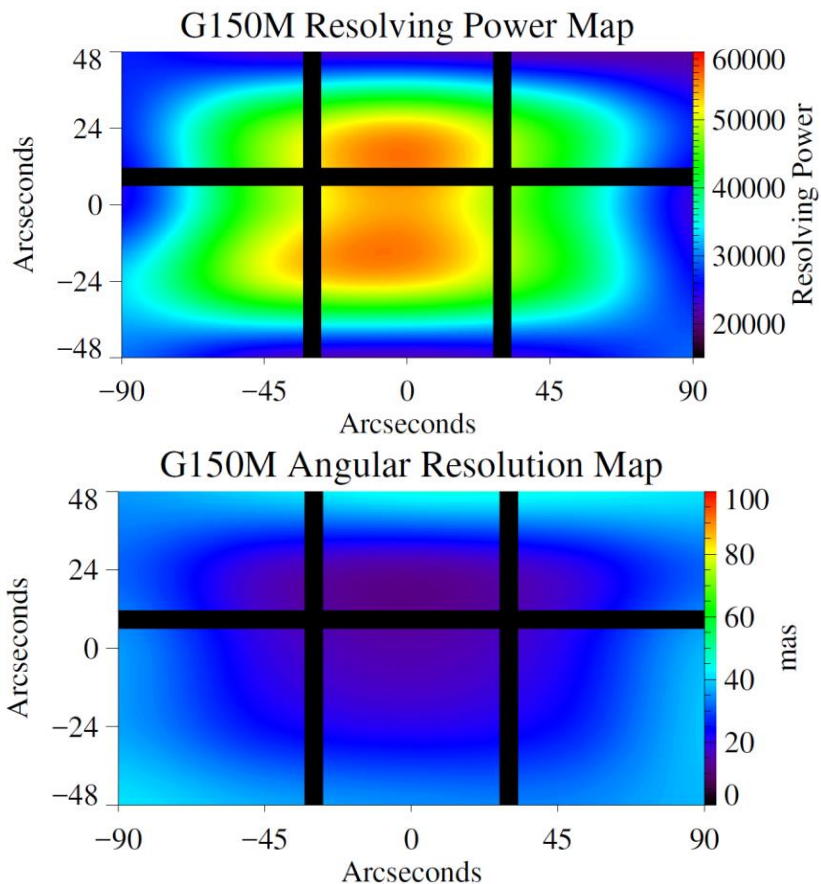

Figure 3 – Spectral resolving power and angular resolution (in milli-arcseconds) maps across the FUV MOS focal plane (using the G150M mode as a representative example). Gaps in the focal plane are created where the MSA arrays are tiled. The center of the OTE focal plane is to the lower left in this representation. The "y" axis is the dispersion direction and the "x" axis is the cross-dispersion direction in these plots.

The ~250 mm diameter medium resolution gratings are toroidal figured and holographically ruled; building on the heritage of HST-COS for aberration-controlling, focusing diffraction gratings. We have discussed the specific grating prescriptions with J-Y Horriba and it has been determined that these gratings could be fabricated on the order of a one-year timescale with existing technology. The observing modes are designed to have roughly 10 nm of optimized overlap with the neighboring setting to allow for robust spectral stitching and allow the user to select the most sensitive or highest angular resolution mode, depending on their science requirements. The actual spectral overlap for an object at the center of the MSA will be ~17 nm, with the additional 7nm having enhanced spectral flaring and lower resolving power. The medium resolution gratings are G120M (optimized for 100 – 140nm; $R$ = 42,000 and angular resolution $\Delta\theta \approx 11$ mas at the center of the field), G150M (130 – 170nm; $R$ = 54,500 and angular resolution $\Delta\theta \approx 15$ mas at the center of the field), and G180M (160 – 200nm; $R$ = 63,200 and angular resolution $\Delta\theta \approx 17$ mas at the center of the field). The center of the FOV is not the optimum field position for these grating modes, however, and are therefore not the peak values. The instrument performance goals and predicted design performance are summarized in Table 1 and Figure 3 shows an example of the spectral and imaging performance of LUMOS G150M mode across the field of view.

The low resolution grating modes are the G155L (100 – 200nm; $R$ = 16,000 and angular resolution $\Delta\theta \approx$ 15 mas at the center of the field) and G145LL (100 – 200nm; $R$ = 500 and angular resolution $\Delta\theta \approx$ 32 mas at the center of the field). The G155L mode, binned up to $R \sim$ 10,000, serves as a good comparison for the only previous medium resolution FUV imaging spectrograph, the STIS G140M mode. LUMOS G155L delivers 20 times the simultaneous bandpass (100nm vs 5nm), six times the one-dimensional field-of-view (180" vs 28"), and more than 100 times the effective area for $\Delta v \sim$ 30 km/s imaging spectroscopy in the FUV. The G145LL mode is designed to minimize detector background and detector footprint, enabling the maximum number of faint objects to be observed simultaneously for FUV spectroscopic deep fields. All of the FUV MOS modes (except G145LL) are focused onto a 2 x 2 array of large-format microchannel plate detectors, with the full spectral bandpass spanning two detector faces, as described in Section 2.3.

*Table 1 -* The instrument parameter goals specified for each mode by the LUMOS instrument team. In the resolving power, bandpass, and angular resolution boxes, the ***target value is on top,*** the ***average value at the center of the field delivered by the LUMOS design is beneath in bold and parentheses***, and the ***average parameter value over 80% of the imaging field-of-view is beneath in bold, italics, and parentheses***. The lower number demonstrates that LUMOS achieves the spectral and spatial resolution goals across the majority of its spectral and spatial detector area.

| *Instrument Parameter* | G120M | G150M | G180M | G155L | G145LL | G300M | FUV Imaging |
|---|---|---|---|---|---|---|---|
| Spectral Resolving Power | 30,000 **(42,000)** *(30,300)* | 30,000 **(54,500)** *(37,750)* | 30,000 **(63,200)** *(40,750)* | 8,000 **(16,000)** *(11,550)* | 500 **(500)** | 30,000 **(40,600)** *(28,000)* | … |
| Optimized Spectral Bandpass (Total) | 100 – 140nm **(92.5 – 147.4 nm)** | 130 – 170nm **(123.4 – 176.6 nm)** | 160 – 200nm **(153.4 – 206.6 nm)** | 100 – 200nm **(92.0 – 208.2 nm)** | 100 – 200nm | 200 – 400nm | 100 – 200nm |
| Angular Resolution | 50 mas **(11 mas)** *(17 mas)* | 50 mas **(15 mas)** *(19.5 mas)* | 50 mas **(17 mas)** *(24 mas)* | 50 mas **(15 mas)** *(27.5 mas)* | 100 mas **(32 mas)** | 50 mas **(8 mas)** *(26 mas)* | 25 mas **(12.6 mas)** *(12.6 mas)* |
| Temporal Resolution | 1 msec | 1 msec | 1 msec | 1 msec | 1 msec | 1 sec | 1 msec |
| Field of View | 2′ × 2′ **(3′ × 1.6′)** | 2′ × 2′ **(3′ × 1.6′)** | 2′ × 2′ **(3′ × 1.6′)** | 2′ × 2′ **(3′ × 1.6′)** | 2′ × 2′ **(3′ × 1.6′)** | 2′ × 2′ **(1.3′ × 1.6′)** | 2′ × 2′ **(2′ × 2′)** |

*NUV Mode* – Undispersed light entering the LUMOS NUV mode follows the same foreoptics (MRM1 and MRM2) before being directed to the G300M grating. The entire NUV wavelength range is accommodated by a single grating, (G300M, 200 – 400nm; $R$ = 40,600 and angular resolution $\Delta\theta \approx$ 8 mas at the center of the field). The dispersed light is then redirected into the NUV detector system by a fixed toroidal fold mirror (NUV MRM3). The NUV detector is a 3 x 7 array of δ-doped CMOS devices that have been derived from the detector systems baselined for the LUMOS HDI imaging system (Bolcar et al. 2017 – this volume; Section 2.3). NUV detector packaging constraints prevent the entire 3′ × 1.6′ field-of-view to be captured on the NUV channel, with a reduction in the cross-dispersion direction field-of-view to 1.3′ × 1.6′.

## 2.2 LUMOS FUV Imaging Mode

The majority of the LUVOIR imaging science is addressed through the HDI instrument (200nm – 2.5μm), and LUMOS will provide a complimentary FUV imaging capability from 100 – 200nm. The LUMOS FUV imaging aperture is physically offset from the MOS MSA array (see Bolcar et al. – this volume), and light from the OTE enters this channel through an unobstructed open aperture. The incident light is folded into the imaging channel off of a biconic convex fold optic with fine piston tip/tilt focus control similar to MRM1 (Imager Relay Mirror 1, IRM1) to a fixed toroidal camera optic (IRM2) and then through two identical reflective filter wheel assemblies (FWA1 and FWA2) that serve to define the imaging bandpass in this mode. A neutral-density filter wheel is inserted between the two filter wheels to accommodate FUV bright object protection, target acquisition for the MOS channel, and safe imaging of the target field through the MSA for shutter selection (see Section 3). The images are then recorded on a single 200mm x 200mm MCP detector (Section 2.3).

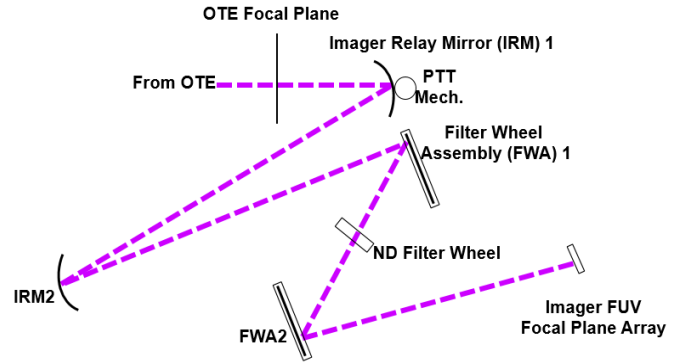

Figure 4 – Schematic raytrace of the LUMOS FUV imaging channel. Light enters through a separate (from the MSA) clear aperture and is folded into the system, imaging bandpasses are defined by reflective multi-layer interference filters in FWA1 and FWA2, before being focused onto the FUV imaging detector, a 200mm x 200mm open-faced MCP employing a cross-strip anode readout. The neutral density (ND) filter wheel is inserted for target acquisition to protect the object against bright objects.

The FUV imaging bands are defined by multi-layer reflective coatings that have heritage from the NUVIEWS rocket[6] and laboratory research and development programs[7,43]. The filter wheels have 7 positions to accommodate 4 "medium band" (FWHM ~ 10nm) filters, one FUV "wide band" (FWHM ~ 30nm) filter, one "open position" (100 – 200nm, where the response is primarily governed by the short wavelength cutoff of LiF on the mirror coatings on the blue end and the detector quantum efficiency, driven by the photocathode, fall-off on the red end) and a "GALEX FUV" filter with an $BaF_2$ filter defines the blue edge and the MCP photocathode cut-off in the red. The "GALEX FUV" filter will also be the band used to image the target field through the MSAs. The 4 medium band filters are: [F110M, F140M, F160M, F180M] with central wavelengths [108nm, 140nm, 160nm, 180nm] and FWHM of approximately 10nm and band-center throughputs of 80% (~55% for the F110M band). The wide band is F150W, centered at 155nm with a 30 nm FWHM and 80% peak throughput. The LUMOS design delivers angular resolution of $\Delta\theta$ = 12.6 mas spots across the entire 2′ × 2′ field of view (see Figure 5).

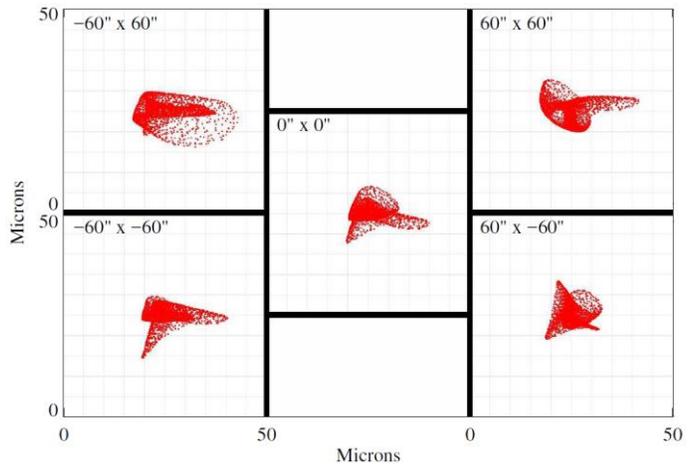

Figure 5. Imaging spot diagram across the 2′ × 2′ field-of-view of the LUMOS FUV imaging channel. The FUV imaging channel is detector-limited to angular resolution < 12 milliarcseconds.

**2.3 LUMOS Optical Coatings and Detectors**

LUMOS optics will utilize protected "enhanced LiF" coatings (eLiF), a high-temperature Al+LiF deposition technique developed as part of a NASA Strategic Astrophysics Technology grant at GSFC[25,27]. These coatings have demonstrated > 85% normal-incidence reflectivity over 1030 – 1300 Å, with sustained performance extending to NIR wavelengths. Testing and flight-qualification of these coatings is an ongoing area of research[27,29]. Advancements in ALD coating technology (e.g., Ref 30) will be leveraged by adding a protective overcoat layer ($AlF_3$ or $MgF_2$) to reduce the hygroscopic sensitivity of the LiF-coated optics. The development and testing of the protected version of eLiF mirrors is being carried out as part of a Nancy Grace Roman Technology Fellowship (RTF), making protected eLiF optics a product of synergies between collaborative NASA SAT, APRA, and RTF efforts. The use of advanced coatings permits the inclusion of the fold and focusing optics without a crippling loss of effective area for the LUMOS imaging and spectroscopy modes.

The LUMOS spectrogram and images are recorded at the focal plane by several large format microchannel plate (MCP) detectors. The use of an MCP device leverages the long heritage of reliable performance that these detectors have with NASA space missions and enables time-tagging capability for the investigation of transient and temporally variable astronomical phenomena (e.g. Ref 20). The imaging mode detector is a single 200mm × 200mm MCP employing a cross-strip anode[31] readout system. The cross-strip readout allows the MCP gain to be dropped by factors of ~10 relative to conventional delay line readouts, providing a correspondingly longer detector lifetime. We have assumed that the 20μm spatial resolution that is being demonstrated with smaller-format cross-strip detectors can be scaled to the large MCP and anode format by the time LUMOS reaches Phase A. The 200mm × 200mm format plates have been demonstrated in the lab[8,16] and will be flight tested in 2017 and 2018 on the NASA-supported DEUCE rocket mission (PI – J. Green, Univ of Colorado). The large format, higher open area afforded by the new ALD borosilicate MCPs (10 – 15% larger open area than conventional MCP glass, with factors of ~3 – 4 lower particle backgrounds; Ref 28), and the optimized pore pitch (up to another 10% efficiency gain) have been enabled by an investment in the NASA-Astrophysics COR technology development. The imaging detector will use an open-face format with an opaque CsI photocathode that has high flight heritage on HST-COS and numerous sounding rocket missions[34,37,45,46].

The LUMOS MOS mode will employ a 2 x 2 array of MCPs, with a 12 mm gap separating the active area of each tile and each tile being the 200mm × 200mm format plates and the cross-strip readout. The MCPs on the blue side of the array will be open-face MCP detectors with CsI photocathodes that have high detector quantum efficiency below 130nm and permit operation down to the short-wavelength edge of LUMOS, 100nm. By design, the small spatial size of the spectral traces of the G145LL mode will be placed exclusively on the open-faced side to ensure full wavelength coverage in every observation. In order to mitigate the long-wavelength losses endemic to the CsI photocathodes, two of the four detector tiles will employ sealed-tube bialkali photocathodes[8,16], a hybrid photocathode material that can increase the response at $\lambda$ > 160nm by factors of 3 – 30 over an open-faced CsI photocathode. The G155L mode uses this enhancement to great effect as the gap in the wavelength coverage between segments is fortuitously placed near the break in the CsI and bialkali quantum response curves.

The increased red sensitivity achieved using the bialkali photocathodes on the red-side MCPs presents certain trades that the observer will want to consider when targeting broad bandpass observations – for instance, the blue end of the G180M band lands on the open-faced CsI MCP, which will have lower sensitivity than the sealed bialkali MCP at those wavelengths (160 – 170nm; CsI has a DQE of ~13% at 165nm whereas the bialkali offers > 30%). In this case, an observer may wish to observe with the G150M setting, where those wavelengths fall on the bialkali detector. All of this would be captured in the eventual LUMOS Instrument Handbook. The MCP detectors are encapsulated in vacuum doors for integration and testing to preserve the integrity of the photocathodes prior to launch. These vacuum doors include a LiF window that permits aliveness testing during integration. These doors would be open for vacuum integration and test, and they would have a one-time opening mechanism that would deploy on-orbit.

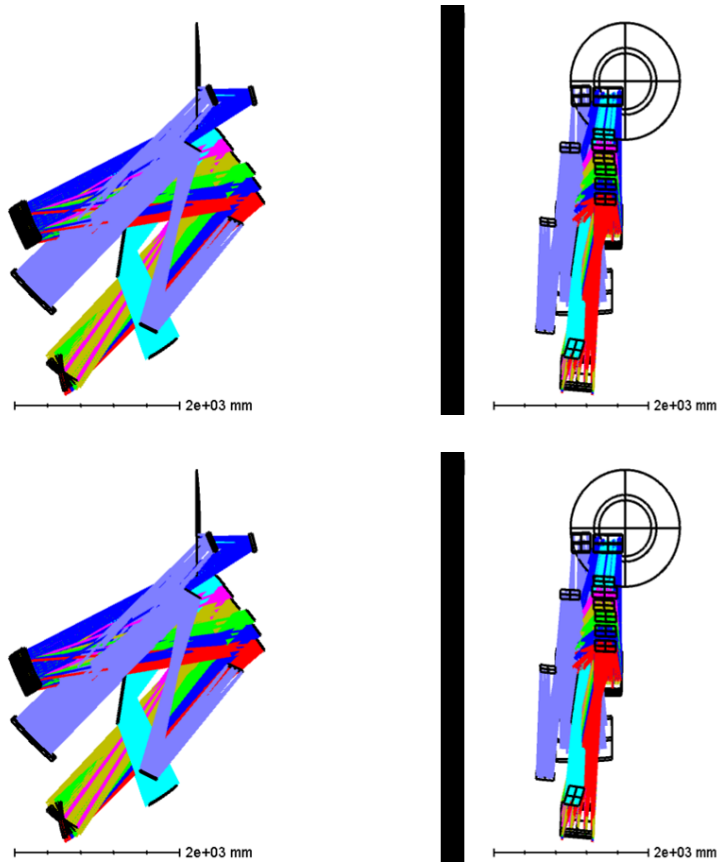

Figure 6 – (above) Raytrace of the LUMOS combined imaging and MOS channels. The representation above left shows a side view while the above right hand view shows LUMOS as observed from behind the OTE primary mirror. The rendering at bottom shows the LUMOS optical path in the mechanical structure.

For the NUV MOS mode, LUMOS will employ a 3 x 7 array of δ-doped CMOS detectors, with 6.5 μm pixel pitch and 8192 x 8192 formats per chip. The FUV performance of δ-doped CCDs[32] is better established (see Refs 33 and 34 for early suborbital results), but the windowing capability and improved radiation hardness led us to select the CMOS detectors for the NUV channel. This choice would be reevaluated at the time of the instrument AO so that the most up-to-date capability and heritage of these detectors is considered when the final LUMOS instrument is defined.

### 3. CONCEPT OF OPERATIONS AND PREDICTED PERFORMANCE

*LUMOS Development and ConOps* - The LUMOS instrument was designed by the UV Astrophysics Group at the University of Colorado's Laboratory for Atmospheric and Space Physics (LASP) and has subsequently undergone a full Instrument Design Laboratory (IDL) run at the Goddard Space Flight Center. During the IDL run, the design was finalized, it was packaged into an optomechanical structure that mounts into the LUMOS instrument bay in the Architecture-A LUVOIR design, instrument mechanisms were specified, a strawman calibration system was designed, data rates and flight software were identified, and instrument performance goals were verified (see Table 1).

Figure 6 shows the final raytrace of the LUMOS design and Figure 1 shows how the optical train is accommodated within the mechanical volume of the LUMOS bay. LUMOS is designed to be serviceable and the LUMOS bay is designed to be removed during instrument upgrades. The LUMOS electronics are fully double-strapped for mechanism and detector redundancy. LUMOS maintains its own data storage capability, enough for 48 hours of high data rate operations (roughly 3 TB). The specification for the MSA shutter failure rate is fewer than 10% of the shutters fail in a five-year mission; the dominant shutter failure mode is 'fail closed'[41,42].

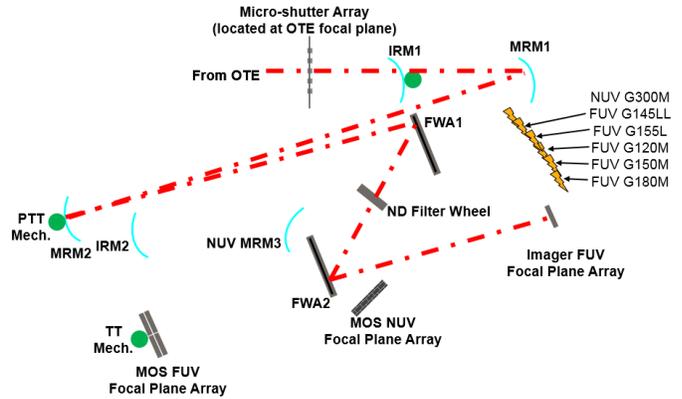

Figure 7 – Schematic raytrace of the LUMOS 'cross-over mode' that is used to directly image the spectroscopic target field through the MSAs for fine-guidance and bright object protection. Undispersed light entering through the MSA is reflected from MRM2 to FWA1 in the imaging channel, though an ND filter wheel and off FWA2, to the FUV imaging detector. This allows target field to be screened for bright objects and for the autonomous target selection and peak-up algorithms to place the spectral targets into the appropriate shutters (Section 2.3).

One long-standing challenge for astrophysics missions employing MCP detectors are limitations on the dynamic range of the detectors and the resultant bright-object protection protocols that are built into flight operations. For example, GALEX avoided most Galactic Plane targets, FUSE was unable to provide significant overlap with the Copernicus hot star sample, and HST-COS had to develop new calibration stars as many of the traditional HST standards were too bright for the sensitive FUV channel. These concerns have been mitigated to a degree by the advent of the cross-strip anodes, which now support global rates as high as several MHz[31] and local rates as high as 100 counts per second per spatial resolution element (O. Siegmund - private communication). Even with these advances, precautions must be taken to avoid potentially damaging overlight situations and longer-term gain sag due to prolonged high illumination rates.

LUMOS performs bright object protection by pre-imaging the target field through the MSA. LUMOS contains a 'cross-over mode' that uses MRM2 to direct light to FWA1 in the imaging channel, though an ND filter wheel, and to the FUV imaging detector. Figure 7 shows a schematic of the cross-over mode, with elements of both the MOS and imaging channels shown. In this way, target brightness is quantified prior to spectral imaging acquisition with the MOS (imaging target acquisitions can also be made through the ND filter for bright-object-protection in the FUV imaging mode). Imaging the spectroscopic target field through the MSA has a second benefit: it enables autonomous microshutter selection and fine-guidance adjustments to acquire the target through the selected shutter/slit. Operationally, the target field is imaged through the MSA, the user-supplied celestial coordinates are references to local detector x-y coordinates, and the instrument autonomously selects the appropriate sources and determines the ΔR.A, ΔDec, and ΔRoll fine adjustments that are required to align the primary science target(s) with the appropriate MSAs and these signals are relayed to the flight computer to complete the target acquisition. After the target has been acquired, a cruciform scan can be performed within an individual shutter to perform a final peak-up prior to beginning science observations.

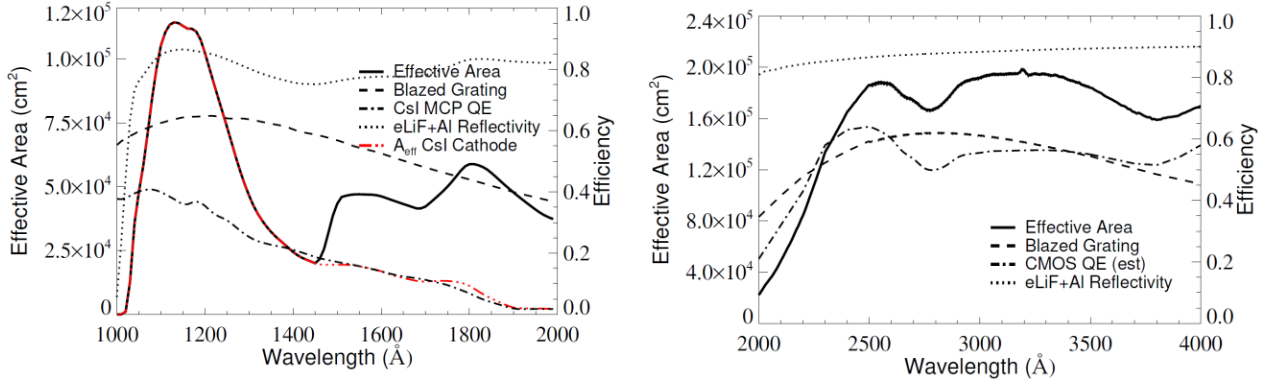

Figure 8 – Effective area of the LUMOS FUV (left) and NUV (right) multi-object spectrograph modes. These values assume the LUMOS Architecture A optical telescope element (15-meter primary diameter) and the component efficiencies described in Section 2.3.

*LUMOS Performance Evaluation* – As part of the initial design, we used the community science documents presented to the LUMOS STDT[47] to develop a set of measurement goals for the LUMOS instrument. At the end of the design phase and in the IDL run, we carried out the detailed modeling required to verify that LUMOS could be built to meet the design goals. The pre-design goals and as-designed specifications are given in Table 1.

We have used the raytrace values to predict the spectral and angular resolution performance for each mode; an example (G150M) is shown in Figure 3. The raytrace has been combined with the total collecting area of the OTE ($A_{geom} = 135 \times 10^4$ cm$^2$), the component efficiencies, and expected background rates to estimate the wavelength-dependent instrumental effective area ($A_{eff}$) and background equivalent flux (BEF). We present the FUV and NUV effective area curves in Figure 8. The effective area, for the FUV MOS, is defined as

$$A_{eff} = A_{geom} * (R_{coat})^7 * G_{grat} * DQE_{det}$$

where $R_{coat}$ is the wavelength-dependent reflectivity of the Al +protected eLiF coating[27], $G_{grat}$ is the wavelength-dependent diffraction grating groove efficiency (taken as the HST-COS values for the FUV modes), and $DQE_{det}$ is the detector quantum efficiency[16,38]. The formulae are modified for the MOS NUV mode and the FUV imager to account for the number of reflections and the lack of the grating in the imaging mode. Taking the expected zodiacal light and interplanetary glow (e.g., spectrally resolved species like H I Lyman-α), combined with the expected detector background rates (estimated to be 1.05 count s$^{-1}$ cm$^{-2}$ for the borosilicate MCPs and 3 electron RMS read-noise and 0.005 counts/hr/pixel of dark rate for the CMOS), we compute the wavelength-dependent BEF in units of [erg s$^{-1}$ cm$^{-2}$ Å$^{-1}$] and plot that in Figure 9. For comparison with the FUV numbers given for LUMOS G155L (which has a field-center resolving power comparable to the COS G130M and G160M modes), the HST-COS M-mode BEF is of order $2 \times 10^{-17}$ erg s$^{-1}$ cm$^{-2}$ Å$^{-1}$ away from strong airglow lines. The lowest BEF for FUV spectroscopy with LUMOS is the G145LL mode, with point source BEF AB-magnitudes between $30 \geq m_{AB} \geq 34$. Figure 10 combines the effective area and BEF results to create a plot of the signal-to-noise per spectral resolution element as a function of source flux for point sources.

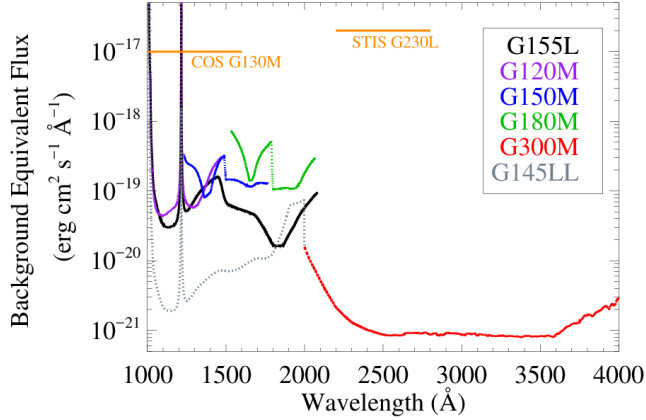

Figure 9 – LUMOS background equivalent flux for the combined FUV and NUV spectroscopic channels. The BEF is driven by detector backgrounds and solar system foregrounds (zodiacal dust and interplanetary atomic emission). The approximate HST limiting spectroscopic BEFs are shown for comparison (orange bar).

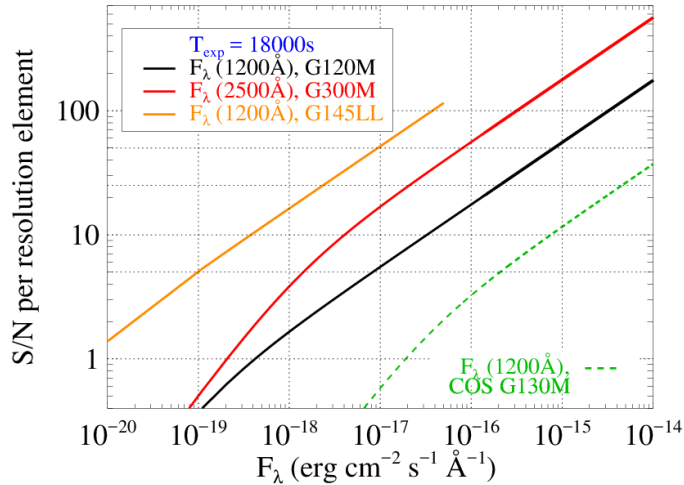

Figure 10 – LUMOS spectroscopic S/N as a function of target flux for a typical long exposure (5 hours). S/N are plotted per resolution element for G120M ($R = 40,000$) at 1200 Å, G300M ($R = 40,000$) at 2500 Å, and G145LL ($R = 500$) at 1200 Å. For comparison, we plot the same values for HST-COS G130M ($R = 17,000$) at 1200 Å.

# 4. ULTRAVIOLET ASTROPHYSICS & PLANETARY SCIENCE WITH THE LUVOIR SURVEYOR

In Section 1, we describe a (very) short tour of the cosmos, highlighting the wide range of astrophysical objects for which spectroscopic observations are essential for determining the physical and chemical state of the system. The common theme among all of these gas-phase diagnostics is that the strongest transitions (determined by their oscillator strengths) of the most abundant species (determined by their relative column densities) reside in the rest-frame ultraviolet bandpass, below the atmospheric cut-off at 3200 Å. Furthermore, with the exception of several million-degree IGM lines (e.g., Ne VIII) which must be redshifted above the Lyman limit at 912 Å and Mg II which is in the near-ultraviolet (NUV) at 2800 Å, all of these lines reside in the far-ultraviolet (FUV) bandpass from approximately 1000 – 1700 Å. For example, O VI ($\lambda\lambda$ 1032, 1038 Å) is a unique tracer of metal-rich gas at a temperature of approximately 300,000 K, HI Ly$\alpha$ (1216 Å) is the strongest line of the most abundant astrophysical species – ubiquitous in almost all astrophysical environments from the cosmic web to Earth's upper atmosphere. The Lyman and Werner ultraviolet band systems of $H_2$ have transition probabilities 15 − 18 *orders of magnitude* larger than the $H_2$ rovibrational 2 − 28 μm) transitions in the near- and mid-infrared, giving us direct access to the dominant molecular reservoir in many dense astrophysical environments without the need to resort to obscure isotopic variants of other molecules with uncertain conversion factors to the total gas mass. Even for the atmospheres of potentially habitable planets, the $O_3$ Hartley bands (2000 – 3000 Å) are by over an order of magnitude the strongest ozone features in a terrestrial atmosphere – particularly useful for Early-Earth-like planets with relatively low atmospheric oxygen abundances. Therefore, the characterization of almost all astrophysical environments requires ultraviolet spectroscopy for a complete quantitative understanding. In the following subsections, we provide several example science investigations that could be carried out as large guest observing programs in a notional 'LUVOIR Cycle 1'. These science programs are presented as '100-hr highlights', programs that would make significant scientific contributions to their sub-fields and astrophysics as a whole with modest investments of observing time on LUVOIR.

## 4.1 100-hour highlight #1: The First 100 Days of Solar System Science with LUMOS

The UV region below 200 nm is relatively free of reflected solar continuum and allows for faint, low surface brightness emissions from fluorescence, impact processes, and recombination to be detected at increased contrast. The most prominent UV features are from giant planet auroral H Ly-$\alpha$ and $H_2$ (Figure 11), which reflects the substantial energy deposition rates in the outer planet ionospheres and the hydrogen dominated atmospheres of the outer planets. In addition, other ion and neutral emissions have been detected from the thin atmospheres of the icy satellites, cometary comae, and planetary exospheres. The extent of our ability to detect and characterize these features has been dominated by the sensitivity of the instruments used and the strength of magnetospheric interactions, local solar wind characteristics, heliocentric distance, and mass loss rates in the various atmospheres under study. Typical airglow features have a surface brightness in the range of ~1000 Rayleighs (R; 1 Rayleigh $\approx$ 80,000 photon $cm^{-2}$ $s^{-1}$ $sr^{-1}$), however spacecraft encounters and Earth orbital remote sensing have identified important features fainter than 10 R (e.g. The Neptune aurora) to as bright as $10^6$ R (e.g. The Jovian aurora). Their characterization is further complicated by temporal variability on time scales from minutes to days, orbit-based perspective changes, and rotational blurring in rapidly-rotating giant planet atmospheres and co-rotating plasma systems.

LUMOS would represent a third revolutionary advance in UV solar system study comparable to those provided by the IUE and HST. With 50-100x the sensitivity of HST and a 25-fold reduction in resolution element (resel) size, LUMOS will be able to detect more confined, rapidly changing or faint emission features (Figure 12), in particular from the ice giant magnetospheres and small bodies in the outer solar system. The wide field and MOS capability of LUMOS will further enable simultaneous spectro-imaging (Figure 13) of interacting magnetospheric structures and of different regions in auroral emission regions. These capabilities combine to provide a set of key science objectives for a '100 day' LUVOIR-LUMOS mission launching in the mid-2030s.

1) Detection and mapping of the auroral emissions from Uranus and Neptune.
2) Simultaneous spectro-imaging of Io-Torus-Auroral Footprint system at Jupiter.
3) Spatial mapping of auroral emissions in the atmospheres of the Galilean satellites.
4) Identification and characterization of plume vents from ocean world satellites of Jupiter.
5) Measurement of outgassing rates on outer solar system bodies (Centaurs, Trojans, Dwarf planets).

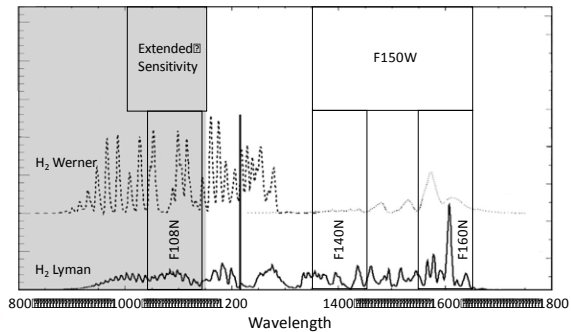

Figure 11. The LUMOS Far-UV sensitivity characteristics are shown relative to the $H_2$ Lyman and Werner band systems detected in outer planet auroras. The expanded wavelength coverage shown at upper left increases sensitivity to the bright Werner bands and therefore the overall detection of auroral emission relative to HST-STIS. The 4 narrow and 1 wide band reflective filters shown allow for full-frame 2D images of emission features. For $H_2$ this would enable direct real-time maps of $CH_4$ absorption (color-ratio) along the aurora ovals.

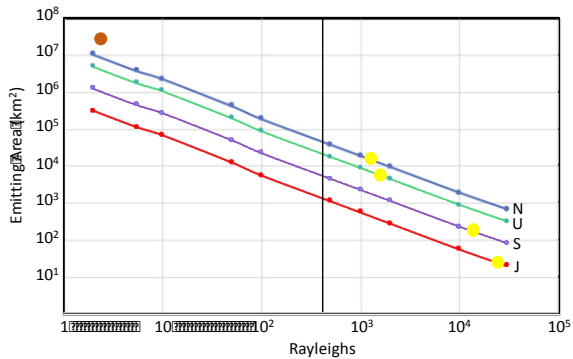

Figure 12. A LUMOS simulation shows the 3s detection limit in a 300 s exposure for extended emission features at the distances of each giant planet systems. The vertical line at ~500 R indicates the physical area (in $km^2$) subtended by a resel at each planet. The yellow circles are the brightness limits for features that can be detected at the $3\sigma$ level before giant planet rotation blurs them beyond the physical size of a resel. The red circle at top left is the measured brightness and area of auroral emission at Neptune as measured by Voyager.

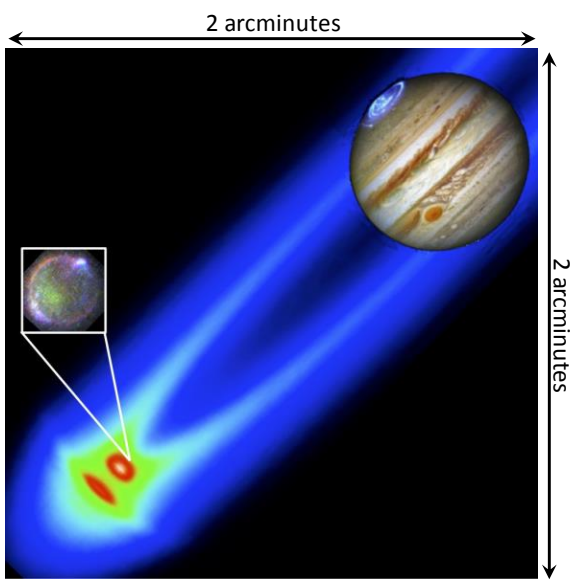

Figure 13. The LUMOS imaging field of view of 2 arcminutes allows for wide field imaging of extended features at 100 mas resolution. Here a full ansa of the Io plasma torus is shown with Jupiter and its aurora. Above the torus is an expanded view of Io, with its own auroral emission shown at the resolution of LUMOS. Simultaneous monitoring of Io, the torus, and both of their interactions with the Jovian auroral system are possible.

## 4.2 100-hour highlight #2: Tracing the composition and evolution of planet-forming material

The Orion complex, at a distance of ~400pc[18], includes star-forming regions subject to different external UV fields and spanning the critical 1-10 Myr age range over which planet-forming disks disperse. As such it represents one of the best sites to study planet formation in action and several of its regions have already a rather complete stellar and disk census (e.g. Fang et al. in prep). Combining 30 times higher sensitivity and 2 - 3 times the spectral resolution than HST-COS with a multi-object imaging spectroscopy mode that enables the simultaneous observation of ~10 - 100 targets (depending on source field density), LUMOS is ideal for ultraviolet spectral surveys of the Orion complex (see

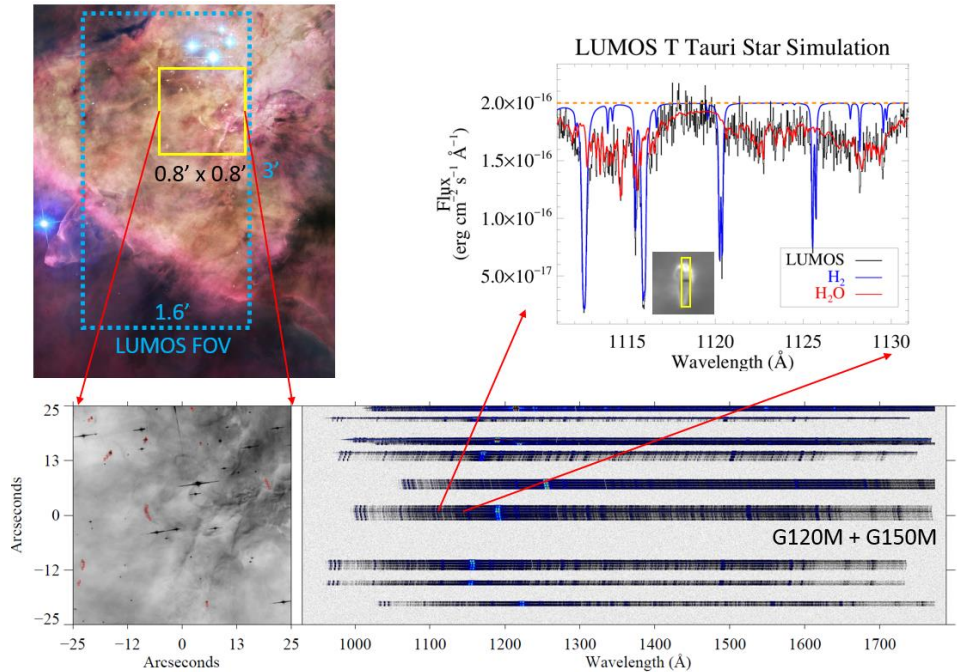

Figure 14 – Multi-object FUV spectroscopy of protoplanetary disks in the Orion Nebula. (*upper left*) HST-ACS image of the Orion Nebula (color credit, Robert Gendler), showing the full FOV of the LUMOS FUV multi-object channel. (*lower panels*) zoom on ~0.8′ × 0.8′ region showing ~30 protostellar/protoplanetary systems[19] and the apertures of the LUMOS microshutter array overplotted (slits are oversized for display). The two-dimensional spectra of the protoplanetary disk and accreting protostar are shown at right. In the *upper right,* we show a zoom in on the 1111 – 1132 Å spectral region containing strong lines of $H_2$ and $H_2O$. The combination of spectral coverage, large collecting area, and multiplexing capability make LUMOS ideal for surveying the composition of the planet-forming environments around young stars.

Figure 14). In about an hour, LUMOS will reach S/N = 10 per resolution element ($R$ = 40,000) over a 3' x 1.6' field to $F_\lambda$= F(1100Å) = 2 x $10^{-16}$ erg $cm^{-2}$ $s^{-1}$ $Å^{-1}$, typical of the flux of young stars in Taurus scaled to the distance of the Orion. This means that in about 2 hours per field LUMOS will cover the entire FUV and NUV spectral range of Orion stars enabling the detection of the most abundant molecular species in disks, e.g. $H_2$, CO, $H_2O$ with contemporaneous measurements of the mass-accretion rate for every star. With pre-defined maps of targets, LUMOS can cover most stars in the Orion Nebula Cluster (ONC; ~1Myr, ~24 square arcminutes size), NGC 1980 (~1-2 Myr, ~16 square arcminutes), σ Ori (~3-5Myr, ~33 square arcminutes), λ Ori (4-8Myr, 49 square arcminutes), 25 Ori (~7-10Myr, 33 square arcminutes) in ~100hr of time at FUV and NUV wavelengths. Such a program will, for the first time, trace the evolution and dispersal of the main molecular carriers of C, H, and O during planet assembly in the terrestrial and giant-planet forming regions. As such it will reveal how the changing disk environment impact the size, location, and composition of planets that form around other stars.

### 4.3 100-hour highlight #3:  Looking Back in Time: The Winds of Metal-Poor Massive Stars

Massive stars are one of the primary engines of the Universe, in life (HII regions, triggered star formation, galactic-scale gas flows), death (long-duration GRBs, core-collapse supernovae) and beyond (neutron stars, black holes, gravitational waves). Many astrophysical simulations ingest the parameterized lifecycle of massive stars – i.e. the duration and properties of their evolutionary stages – to account for their ionizing, mechanical, and chemical feedback on their local environments. However, significant gaps in our knowledge remain. **A top priority of the field is to characterize massive stars at very low metallicities (0.1-0.01 solar).** This regime is crucial to interpret galaxies at intermediate/high redshifts, high-mass transients (GRBs, SNe, GW sources) and, most importantly, to anchor the physics of the (near) metal-free First Stars that shaped the early Universe.

The largest difference between two stars born with the same mass but different metallicities is the physics of their winds, outflows of mass powered by absorption of photons by metals in the atmosphere. The amount of mass the star loses via the wind at the hot stages (O-B2 types, Wolf-Rayet, LBV) depends on metallicity, with dramatic consequences on stellar evolution and the size of the pre-SN cores. The winds of metal-poor massive stars are ill-constrained at present, and this is largely due to a lack of suitable UV observations.

With 100hr of observing time with LUVOIR-LUMOS, we would observe 300 metal-poor massive stars in environments of decreasing metallicity (see Table), enabling:

- Definitive characterization of metal-poor winds, including the effects of shocks, inhomogeneities, and the **true mass-loss rates**;
- Study the metallicity dependence of stellar winds in the metal-poor (sub-SMC) regime for the first time, giving a **parameterization of mass-loss with metallicity that could be extrapolated to the First Stars.**

Our observations require the proposed G155L setting to fully resolve the wind structures and P Cygni profiles, and a multiplex of ≥10. We note that I Zw18 does not have a catalog of resolved targets yet, but this is expected in the coming years (e.g. *JWST,* TMT). Adequate candidate stars are already known in the other target galaxies, but current UV facilities simply lack the sensitivity and multiplex required to assemble the observational samples needed to provide empirical calibration and testing of the evolutionary models. Moreover, these catalogs will likely be improved as well, providing new candidates. Estimated targets densities are provided in the Table.

The project requires SNR≥20 per resolution element at 1500Å. According to the online ETC this is exceeded for the O-dwarfs of IC1613 in 0.5 hours regardless the final aperture of LUVOIR, and 1h for O-stars at 1.5 Mpc. LUMOS will reach the brightest stars of IZw18 in 11.5h with the 15.1m aperture. (Observations with the 9.2m aperture would still be feasible but binning the data to R~2500). Several pointings per galaxy are needed to both cover all interesting locations and a substantial percentage of the targets.

| Galaxy | Distance / Proj. Diam | Metallicity | #Known OB(WR+LBV) Expected density | GALEX FUV $m_{AB}$ | #Pointings | $T_{exp}$ per field |
|---|---|---|---|---|---|---|
| IC 1613 | 750 Kpc 13' | 1/7 Osun | 56 (2) 54 stars/arcmin$^2$ (¤) | 16.69 B0I 19.19 OV | 8 | 0.5h |
| Sext-A, WLM, NGC3109 | ≤1.5 Mpc 5', 8.5', 14' | 1/10 Osun | 12 (-), 8 (-), 44 (-) 74 stars/arcmin$^2$ (¤) | 18.19 20.69 | 3 x 5 | 1h |
| I Zw18 | 18 Mpc 30" | 1/32 Osun | – 2515 stars/arcmin$^2$ (†) | 23.70 26.20 | 7 | 11.5h |
| **TOTAL: 300 stars** | | | | | 30 | ~100h |

(¤) Estimated with color-cuts on ground-based photometry. (†) Estimated with the number of known OB stars in 30 Doradus that would be bright enough in IZw18 to be observed with LUVOIR ($M_V$=-6.0).

### 4.4 100-hour highlight #4: Galaxy Fueling and Quenching in the Low and Intermediate Redshift Universe

The gas flows that drive galaxy accretion and feedback are critical, but still poorly understood, processes in their formation and evolution. One of Hubble's successes has been in characterizing the CGM that spans 30 times the radius and 10000 times the volume of the visible stellar disk. Thanks to Hubble and its ground-based optical collaborators, we know roughly how much matter the CGM contains, but the extremely low densities make it difficult to ascertain its exact role in galaxy evolution. The critical question is how this gas enters and leaves the galaxies: galactic star formation rates are limited by the rate at which they can acquire gas from their surroundings, and the rate at which they accumulate heavy elements is limited by how much they eject in outflows. Much of the still-unknown story of how galaxies formed comes down to how they acquire, process, and recycle their gas.

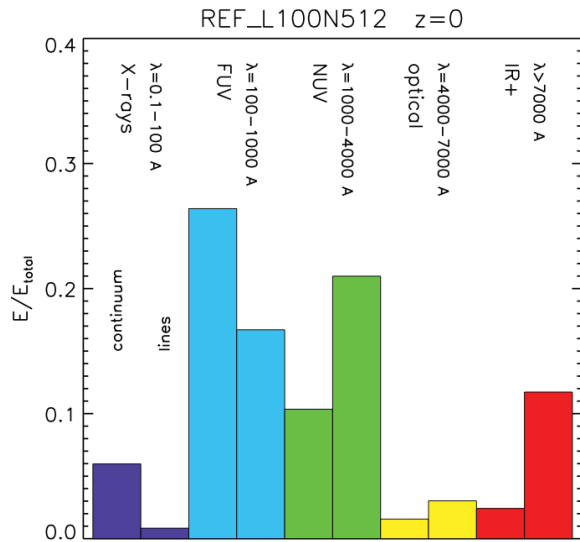

Hubble can make only crude statistical maps by sampling many halos with one absorption-line path through each. The future challenge is to "take a picture" of the CGM using the LUMOS G155L mode. This is intrinsically a UV problem since most of the energy transferred by diffuse gas on its way in or out of galaxies is emitted or absorbed in rest-frame UV lines of H, C, O, Ne, and other metals, including rest-frame extreme-UV lines that redshift into the 1000-2000 Å band at $z > 0.5$ (Ref 49; Fig. 15). In 100 hours of observing time, LUMOS will be able to map the density, temperature, and mass flow rates of the CGM, directly, using the UV radiation emitted by CGM gas as it cycles in and out of galaxies. Observing up to 50-100 sources at a time, LUVOIR could map the faint light (e.g., an O VI halo at $z \sim 0.2$ with $S_B > 300$ photons cm$^{-2}$ s$^{-1}$ sr$^{-1}$ to 3σ) emitted by gas entering and leaving galaxies, to count up the heavy element content of this gas, to watch the flows as they are ejected and recycled, and to witness their fate when galaxies quench their star formation, all as a function of galaxy type and evolutionary state.

**Figure 15:** The bars show the fraction of energy emitted in various rest-frame bands by gas cooling radiation in a hydrodynamical galaxy formation simulation[49]. More than half of the radiation is emitted in the rest-frame FUV (100-1000 Å) and most of the remainder is in the NUV (1000-4000 Å).

### 4.5 100-hour highlight #5: Lyman Continuum Luminosity Function Evolution $0.1 < z < 1.2$

During the epoch of reionization (EOR), somewhere between a redshift of $12 > z > 6$[51,53], the universe underwent a phase change driven by ionizing radiation (Lyman continuum - LyC) emitted by the first collapsed objects at wavelengths below the hydrogen ionization edge at 911.8 Å. This emission transformed the universe from a radiation-bounded topology, wherein budding H II regions were embedded in primordial neutral media, into to a density-bounded topology, wherein the neutral material became enveloped in an expanding meta-galactic ionizing background (MIB). The fraction of ionizing radiation, $f^e_{LyC}$, escaping from these first and subsequent objects played a crucial role in regulating the emergence and sustenance of the MIB and the concomitant evolution of large-scale structures over cosmic timescales.

It is widely acknowledged that $f^e_{LyC}$ is an unknown. It represents a major systematic uncertainty[52,54] not only for understanding the EOR on a hierarchical scale, but also for understanding the coupling of ionizing radiation to the secular evolution of galaxies from $6 > z > 0$. Whether the MIB was primarily created by ionizing radiation emitted by the first black holes or stars is a matter of debate, however, it is commonly estimated that small galaxies dominated the ionizing radiation budget, provided the faint end slope of the galaxy luminosity function was steep and extended to absolute UV magnitudes $M_{1500} < -13$[54]. A key project for *JWST* is to determine the source(s) responsible for the EOR transition. Unfortunately, *JWST* cannot directly observe $f^e_{LyC}$ because the LyC from high $z$ is attenuated along our line-of-sight by overlapping clouds of neutral hydrogen with column densities $\log(N_{HI}(cm^2)) > 17.2$. Estimates for the mean transmission at $(1+z)911.8$ Å through the IGM from objects at redshifts of z = [0.5, 1, 2, 3, 4, 5, 6] are $T_{IGM}$ = [0.97, 0.96, 0.8, 0.5, 0.3, 0.08, 0.01][55,56]. The UV has a clear advantage of only requiring trivial corrections for IGM attenuation in efforts to assess the environmental conditions that favor ionizing radiation escape, and to test whether faint star-forming galaxies analogous to those at high redshift could have driven the EOR.

In previous work, McCandliss & O'Meara (2017)[56] provided requirements for the detection of LyC escape from star-forming galaxies for $z < 3$ as derived from the Arnouts et al. (2005) UV luminosity functions[50]. Here we use those detection requirements for objects in the redshift interval from $0 < z < 1.2$ along with the LUMOS effective area and background equivalent flux (*BEF*), to assess science return. Our metric is the total number of objects in the FOV = $3.'0 \times 1.'6$, over a given redshift interval that can be observed in 10 hours to $f^e_{900}$ = [1.00, 0.64, 0.32, 0.16, 0.08, 0.04, 0.02, 0.01] at 5σ confidence, assuming a bandpass of $\Delta\lambda$ = 30 Å, and flux limit of 3*BEF*, using the "LowLow" resolution mode of LUMOS (FUV G145LL, $R \approx 500$). In the left panel of Figure 1 we show the cumulative number of galaxies that meet these criteria. The results summarized in Table 1, showing 1028 objects detectable if they have $f^e_{900}$ = 1; 786 objects if $f^e_{900}$ = 0.64; 525 objects if $f^e_{900}$ = 0.32... This is a good match to the number of lines in the MSA Assembly (MSAA).

In the right panel of Figure 16 we show the absolute 1500 Å magnitudes of the objects probed. We conclude that **LUMOS can easily reach those objects at the faint end of the luminosity function (-14 < $M_{1500}$ < -8) that are analogous to the star-forming galaxies thought to be responsible of reionization.** In a one-hundred hour observing program there will be ten times the number of objects shown in Table 1, which will be sufficient to establish the levels of LyC escape and to begin a statistically significant assessment of LyC luminosity function evolution over these redshift intervals.

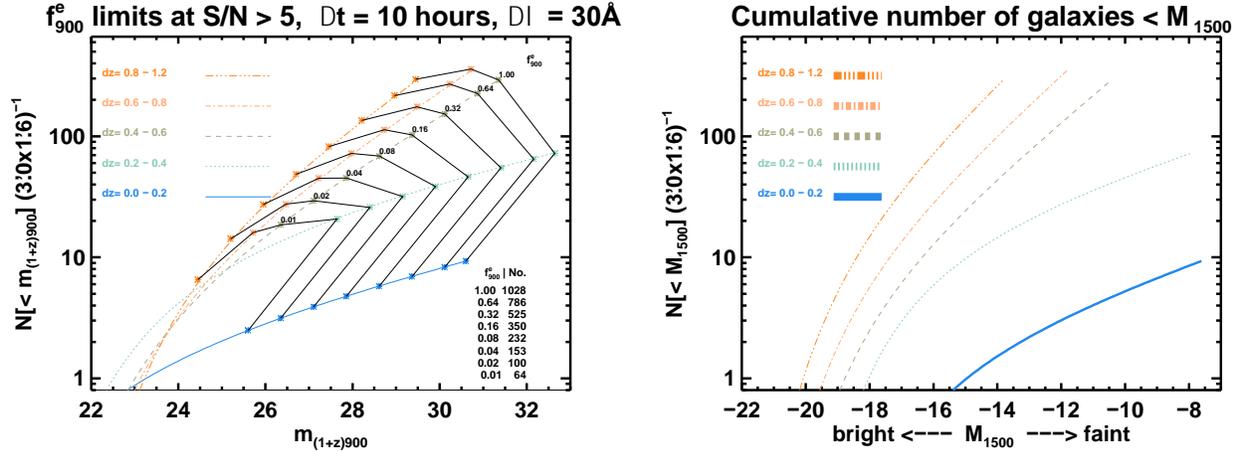

Figure 16. Right – Cumulative number of galaxies in the LUMOS FOV brighter than the apparent 900 Å magnitude ($m_{(1+z)900}$) derived from the differential luminosity functions[50]. See text for assumptions. Left – Cumulative number of galaxies brighter than the absolute 1500 Å magnitude (M1500) to the $f^e_{900}$ =1 escape limit.

|  | N(Δz) | | | | | Totals |
|---|---|---|---|---|---|---|
| $f^e_{900}$ | Δz = (0 – 0.2) | Δz = (0.2 – 0.4) | Δz = (0.4 – 0.6) | Δz = (0.6 – 0.8) | Δz = (0.8 – 1.2) | |
| 1.00 | 9 | 72 | 292 | 358 | 295 | 1028 |
| 0.64 | 8 | 64 | 225 | 270 | 217 | 786 |
| 0.32 | 6 | 54 | 152 | 175 | 135 | 525 |
| 0.16 | 5 | 46 | 102 | 113 | 82 | 350 |
| 0.08 | 4 | 38 | 68 | 72 | 48 | 232 |
| 0.04 | 3 | 31 | 45 | 45 | 27 | 153 |
| 0.02 | 3 | 25 | 29 | 27 | 14 | 100 |
| 0.01 | 2 | 20 | 18 | 15 | 6 | 64 |

Table 2 Number of observable galaxies in FOV = 3.′0 × 1.′6 as a function of redshift interval Δz to indicated $f_{esc}$ limit, 5σ, Δt = 10 hours, Δλ = 30.

## 5. SUMMARY

We have presented the first description of a multiplexed UV spectrograph for NASA's LUVOIR Surveyor mission concept, the *LUVOIR Ultraviolet Multi-Object Spectrograph*. LUMOS offers point source and multi-object spectroscopy across the UV bandpass, with multiple resolution modes to support different science goals. The instrument will provide low ($R = 8,000 – 18,000$) and medium ($R = 30,000 – 65,000$) resolution modes across the far-ultraviolet (FUV: 100 – 200 nm) and near-ultraviolet (NUV: 200 – 400 nm) windows, and a very low resolution mode ($R = 500$) for spectroscopic investigations of extremely faint objects in the FUV. Imaging spectroscopy will be accomplished over a 3 × 1.6 arcminute field-of-view by employing a holographically-ruled diffraction grating to control optical aberrations, microshutter arrays (MSA), advanced optical coatings for high-throughput in the FUV, and next generation large-format photon-counting detectors. The spectroscopic capabilities of LUMOS are augmented by an FUV imaging channel (100 – 200nm, 13 milliarcsecond angular resolution, 2 × 2 arcminute field-of-view) that will employ a complement of narrow- and medium-band filters. We describe the instrument requirements for LUMOS, the instrument design, and technology development recommendations to support the hardware required for LUMOS. We present an overview of LUMOS' observing modes and estimated performance curves for effective area, spectral resolution, and imaging performance. Example "LUMOS 100-hour Highlights" observing programs are presented to demonstrate the potential power of *LUVOIR*'s ultraviolet spectroscopic capabilities.

*Acknowledgements* - This work was supported in part by NASA grants NNX13AF55G and NNX16AG28G to the University of Colorado at Boulder. The authors thank Drs. Manuel Quijada, John Vallerga, and Oswald Siegmund for helpful feedback regarding detector and coating capabilities for the next decade. The authors also acknowledge enjoyable discussions with the LUVOIR STDT in developing the science requirements for the LUMOS instrument and the Goddard Space Flight Center Instrument Design Lab for technical assistance during the final design and implementation phases.

**REFERENCES**

[1] 2010 Decadal Survey, "New Worlds, New Horizon's in Astronomy and Astrophysics", National Academies Press, 2010
[2] France, Kevin, "CHISL: the combined high-resolution and imaging spectrograph for the LUVOIR surveyors", JATIS, v2, 1203, 2016
[3] NASA, "2013 Astrophysics Roadmap: Enduring Quests, Daring Visions: NASA Astrophysics in the next Three Decades", 2013
[4] Seager, S., et al., "From Cosmic Birth to Living Earth: A visionary space telescope for UV-Optical-NearIR Astronomy", AURA, http://www.hdstvision.org/report/, 2015
[5] Sembach, K., et al. "Cosmic Origins Program Analysis Group (COPAG) Report to Paul Hertz Regarding Large Mission Concepts to Study for the 2020 Decadal Survey", NASA COPAG, http://cor.gsfc.nasa.gov/copag/rfi/, 2015
[6] Park, J., Kim, J., Zukic, M., Torr, D. G. 1994, "Reflective filters design for self-filtering narrowband ultraviolet imaging experiment wide-field surveys (NUVIEWS) project", Proc. SPIE 2279:155
[7] Rodríguez-De Marcos, Luis; Larruquert, Juan I.; Méndez, José A.; Aznárez, José A.; Vidal-Dasilva, Manuela; Fu, Liping "Narrowband filters for the FUV range", Proc. SPIE, Volume 9144, 2014
[8] Siegmund, O.; Vallerga, J.; Tremsin, A.; McPhate, J.; Frisch, H.; Elam, J.; Mane, A.; Wagner, R.; Varner, G.. "Large Area and High Efficiency Photon Counting Imaging Detectors with High Time and Spatial Resolution for Night Time Sensing and Astronomy", Proceedings of the Advanced Maui Optical and Space Surveillance Technologies Conference, 2012
[9] Roberge, A., et al. "FUSE and Hubble Space Telescope/STIS Observations of Hot and Cold Gas in the AB Aurigae System", ApJL, 551, 97, 2001
[10] France, Kevin, et al. "CO and $H_2$ Absorption in the AA Tauri Circumstellar Disk", ApJ, 743, 186, 2012c
[11] France, Kevin; Herczeg, Gregory J.; McJunkin, Matthew; Penton, Steven V. "$CO/H_2$ Abundance Ratio ≈ $10^{-4}$ in a Protoplanetary Disk", ApJ, 794, 160, 2014
[12] Roberge, A., et al. "High-Resolution Hubble Space Telescope STIS Spectra of C I and CO in the β Pictoris Circumstellar Disk", ApJ, 538, 904, 2000
[13] France, Kevin, et al. "The Far-ultraviolet "Continuum" in Protoplanetary Disk Systems. II. Carbon Monoxide Fourth Positive Emission and Absorption", ApJ, 734, 31, 2011
[14] France, Kevin, et al. "A Hubble Space Telescope Survey of $H_2$ Emission in the Circumstellar Environments of Young Stars", ApJ, 756, 171, 2012
[15] Seager, S.; Bains, W.; Hu, R., "Biosignature Gases in $H_2$-dominated Atmospheres on Rocky Exoplanets", APJ, 777, 95, 2013

[16] Ertley, C. D.; Siegmund, O. H. W.; Jelinsky, S. R.; Tedesco, J.; Minot, M. J.; O'Mahony, A.; Craven, C. A.; Popecki, M.; Lyashenko, A. V.; Foley, M. R.. "Second generation large area microchannel plate flat panel phototubes", Proc. SPIE, Volume 9915, 2016
[17] Tian, Feng, et al., 2014, "High stellar FUV/NUV ratio and oxygen contents in the atmospheres of potentially habitable planets", E&PSL, 385, 22
[18] Menten, K. M.; Reid, M. J.; Forbrich, J.; Brunthaler, A., "The distance to the Orion Nebula", A&A, 474, 515, 2007
[19] Bally, John; Sutherland, Ralph S.; Devine, David; Johnstone, Doug., "Externally Illuminated Young Stellar Environments in the Orion Nebula: Hubble Space Telescope Planetary Camera and Ultraviolet Observations", AJ, 116, 293, 1998
[20] France, Kevin, et al. "The MUSCLES Treasury Survey: Motivation and Overview", ApJ, *submitted*, 2016b
[21] Des Marais, David J., et al., "Remote Sensing of Planetary Properties and Biosignatures on Extrasolar Terrestrial Planets", AsBio, 2, 153, 2002
[22] Kaltenegger, Lisa, Traub, Wesley A., & Jucks, Kenneth W., "Spectral Evolution of an Earth-like Planet", ApJ, 658, 598, 2007
[23] Seager, S., Deming, D., & Valenti, J. A., "Transiting Exoplanets with JWST", Astrophysics in the Next Decade, Astrophysics and Space Science Proceedings, Springer, 123, 2009
[24] Bétrémieux, Y. & Kaltenegger, L., "Transmission Spectrum of Earth as a Transiting Exoplanet from the Ultraviolet to the Near-infrared", ApJ, 772, 31, 2013
[25] Quijada, M. et al. "Enhanced far-ultraviolet reflectance of $MgF_2$ and LiF over-coated Al mirrors", SPIE, v9144, 2014
[26] Balasubramanian, Kunjithapatham, et al., "Aluminum mirror coatings for UVOIR telescope optics including the far UV", SPIE, 9602, 01, 2015
[27] Fleming, Brian T., et al., "New UV instrumentation enabled by enhanced broadband reflectivity lithium fluoride coatings", 9601, 0R, 2015
[28] Siegmund, O. H. W, et al., "Advances in microchannel plates and photocathodes for ultraviolet photon counting detectors", SPIE, 8145, 0J, 2011
[29] France, Kevin, et al. "The SLICE, CHESS, and SISTINE Ultraviolet Spectrographs: Rocket-borne Instrumentation Supporting Future Astrophysics Missions", JAI, *in press*, 2016
[30] Moore, Christopher S.; Hennessy, John; Jewell, April D.; Nikzad, Shouleh; France, Kevin,"Recent developments and results of new ultraviolet reflective mirror coatings", SPIE, 9144, 4, 2014
[31] Vallerga, John, et al. "Cross strip anode readouts for large format, photon counting microchannel plate detectors: developing flight qualified prototypes of the detector and electronics", SPIE, 9144, 3, 2014
[32] Nikzad, Shouleh, et al., "Delta-doped electron-multiplied CCD with absolute quantum efficiency over 50% in the near to far ultraviolet range for single photon counting applications", ApOpt, 51, 365, 2012
[33] France, K.; Andersson, B.-G.; McCandliss, S. R.; Feldman, P. D., "Fluorescent Molecular Hydrogen Emission in IC 63: FUSE, Hopkins Ultraviolet Telescope, and Rocket Observations", ApJ, 628, 750, 2005
[34] Lupu, Roxana E.; McCandliss, Stephan R.; Fleming, Brian; France, Kevin; Feldman, Paul D.; Nikzad, Shouleh "Calibration and flight performance of the long-slit imaging dual order spectrograph", SPIE, 7011, 2008
[35] France, Kevin; Beasley, Matthew; Kane, Robert; Nell, Nicholas; Burgh, Eric B.; Green, James C. "Development of the Colorado High-resolution Echelle Stellar Spectrograph (CHESS)", 8443, 05, 2012
[36] France, K. et al. "Flight performance and first results from the sub-orbital local interstellar cloud experiment (SLICE)", SPIE, 8859, 10, 2013b
[37] Hoadley, K., France, K. et al. "The re-flight of the Colorado high-resolution Echelle stellar spectrograph (CHESS): improvements, calibrations, and post-flight results", SPIE, v9905, 2016
[38] Siegmund, Oswald H. W.; Tremsin, Anton S.; Vallerga, John V. "Development of cross strip MCP detectors for UV and optical instruments" SPIE, 7435, 0L, 2009
[39] Ake, T. B. et al. "COS FUV Flat Fields and Signal-to-Noise Characteristics", HST Calibration Workshop, 2010
[40] Harman, C. E.; Schwieterman, E. W.; Schottelkotte, J. C.; Kasting, J. F. "Abiotic $O_2$ Levels on Planets around F, G, K, and M Stars: Possible False Positives for Life?", ApJ, 812, 137, 2015
[41] Li, M. J.; Brown, A. D.; Kutyrev, A. S.; Moseley, H. S.; Mikula, V. "JWST microshutter array system and beyond", SPIE, 7594, 75940, 2010
[42] Kutyrev, A. S.; Collins, N.; Chambers, J.; Moseley, S. H.; Rapchun, D. "Microshutter arrays: high contrast programmable field masks for JWST NIRSpec", SPIE, 7010, 70103, 2008
[43] Zukic, M., Torr, D. G., Kim, J., Spann, J. F., and Torr, M. R., "Far ultraviolet filters for the

ISTP UV imager", Opt. Eng. 32:3069

[43] France., K.; Herczeg, G., McJunkin, M., and Penton, S. "CO/$H_2$ Abundance Ratio Observed in a Protoplanetary Disk", ApJ, 794, 160, 2014

[44] Carr, J. & Najita, J. "Organic Molecules and Water in the Inner Regions of T Tauri Stars", ApJ, 733, 102, 2011

[45] Green, James C.; Froning, Cynthia S., et al. "The Cosmic Origins Spectrograph", ApJ, 744, 60, 2012

[46] Fleming, Brian T; McCandliss, Stephan R.; et al. "Fabrication and calibration of FORTIS", SPIE 8145, 2011

[47] France, Kevin "The LUVOIR science and technology definition team (STDT): overview and status", SPIE 9904, 2016

[48] Nikzad, Shouleh; Jewell, April D.; et al. "High Efficiency UV/Optical/NIR Detectors for Large Aperture Telescopes and UV Explorer Missions: Development of and Field Observations with Delta-doped Arrays", JATIS- in press, 2017

[49] Bertone, Serena; Aguirre, Anthony; Schaye, Joop, "How the diffuse Universe cools", MNRAS, 430, 3292, 2013

[50] Arnouts, S.; Schiminovich, D.; Ilbert, O. "The GALEX VIMOS-VLT Deep Survey Measurement of the Evolution of the 1500 Å Luminosity Function", ApJ, 619, 43, 2005

[51] Bouwens, R. J., et al., "Reionization After Planck: The Derived Growth of the Cosmic Ionizing Emissivity Now Matches the Growth of the Galaxy UV Luminosity Density", ApJ 811, 140, 2015

[52] Ellis, R. S., "Cosmic Dawn: Studies of the Earliest Galaxies and Their Role in Cosmic Reionization", arXiv:1411.3330, 2014

[53] Fan, X., C. L. Carilli, and B. Keating, "Observational Constraints on Cosmic Reionization" ARA&A44, 415–462, 2006

[54] Finkelstein, S. L., et al., "The Evolution of the Galaxy Rest-frame Ultraviolet Luminosity Function over the First Two Billion Years", ApJ 810, 71, 2015

[55] Inoue, A. K., I. Shimizu, I. Iwata, and M. Tanaka, "An updated analytic model for attenuation by the intergalactic medium", MNRAS 442, 1805–1820, 2014

[56] McCandliss, S. R. and J. M. O'Meara, " Flux sensitivity requirements for the detection of Lyman continuum radiation from star-forming galaxies below redshifts of 3", Submitted ApJ 2017